# Localized orientational order chaperons the nucleation of Rotator phases in hard polyhedral particles

(Nov. 8, 2013 version)

Vikram Thapar and Fernando A. Escobedo
School of Chemical and Biomolecular Engineering, Cornell University, Ithaca, NY14853

## Abstract

The nucleation kinetics of the rotator phase in hard cuboctahedra, truncated octahedra, and rhombic dodecahedra is simulated via a combination of Forward Flux Sampling and Umbrella Sampling. For comparable degree of supersaturation, the polyhedra are found to have significantly lower free-energy barriers and faster nucleation rates than hard spheres. This difference primarily stems from localized orientational ordering, which steers polyhedral particles to pack more efficiently. Orientational order hence fosters here the growth of orientationally disordered nuclei.#

Homogeneous nucleation is the key step in spontaneous crystallization in which an embryo of a stable solid-phase is created within a metastable liquid. Understanding nucleation remains a challenge, as it is an activated process that involves overcoming a free-energy barrier by a rare fluctuation in the metastable phase. Conceptually, classical nucleation theory (CNT) [1] is widely used to provide a framework to study nucleation kinetics. CNT assumes that each nucleus is spherical and formed by the single most thermodynamically stable solid. Consequently, the size of the largest nucleus serves as the only relevant order parameter to monitor the kinetics of the process whose transition state defines the critical nucleus size. Although useful, this simplified picture has also proven to be incomplete. For Lennard-Jones spheres, e.g., Moroni *et al.* [2] showed that not only size but also the shape and structure of the critical nuclei are important to describe the nucleation mechanism. Other known limitations of CNT are that it applies bulk thermodynamic properties to clusters comprising only $O(10^2)$ particles and that it cannot be used to explain multi-stage nucleation.

Molecular simulations provide a more rigorous tool to probe the mechanism of nucleation. However, conventional brute force simulations are impractical as nucleation is a rare event, which makes it difficult to collect enough statistics with computing resources typically available. To overcome this limitation, many sophisticated methodologies have been developed [3-12]. Amongst them, Forward Flux Sampling (FFS) [11,12] is particularly appealing for studying nucleation given its ability to simultaneously and efficiently resolve the transition path ensemble (TPE), the committor probabilities, the transition state ensemble (TSE) and the transition rates. FFS generates the TPE by "ratcheting" partial trajectories forward (and over large free-energy barriers) by the use of interfaces (as steps on a ladder) created along an order parameter that tracks the phase space between the two basins of interest.

Despite the multiple simulation studies and different techniques that have been used to examine nucleation processes, most of them have been concerned with spherical particles given their simplicity and relevance to some atomic and colloidal systems. At present, however, a major focus in material science is the self-assembly of anisotropic particles due to its potential to help engineer novel materials from colloidal nanoparticles [13,14]. Assemblies of anisotropic particles undergo order-disorder phase transitions involving changes in both translational and orientational degrees of freedom and can lead to phases with partial structural order or "mesophases" [15]. Recent studies [16-19] have provided a partial roadmap of the mesophases that can be expected for different types of polyhedral particles. Along with the phase behavior, understanding the kinetics of the phase transition of anisotropic particles is also of practical importance (e.g., kinetic traps may preclude the timely formation of a desirable phase) and yet this has been either scantly explored (for rod-like particles) [20-23] or completely unexplored (for polyhedral particles).

In this work, we focus on the rotator-phase nucleation in hard cuboctahedra (CO), truncated octahedra (TO), and rhombic dodecahedra (RD). They are representative of a class of polyhedral particles that form crystals with well-defined translational and orientational order at high concentrations [16,18,19]. At the order-disorder-transition pressure (ODP), the isotropic phase of suspensions of these polyhedra transitions into a mesophase that exhibits a *rotator* or *plastic* character; in such a state the particles have translational order but are essentially orientationally disordered. Given that primarily translational order is nucleated near their ODP, one could conjecture that the kinetics of the rotator-phase nucleation of COs, TOs, and RDs is comparable to that for the nucleation of translational order in suspensions of hard-spheres (HSs). This is particularly relevant in comparing COs and HSs since, as it is shown in the *Supplemental Material* [24], the rotator phase in COs has (unlike those for TOs and RDs) minimal long-range orientational order at all pressures where it is stable. We show here that the above conjecture of kinetics similarity is incorrect. We find that while the size, shape, and structure of the sub-critical and critical solid nuclei do bear similarities to those seen in HSs, for COs, TOs, and RDs the free-energy barrier is significantly lower, and the nucleation rate is significantly faster than those in HSs for comparable degrees of supersaturation. We identify small regions of orientationally ordered particles likely acting as *catalyst* for the rotator-phase nucleation in the polyhedra (such regions are local fluctuations that foreshadow the establishment of long-range orientational order to ensue at higher pressures).

FFS has been used to study the crystal-nucleation in various systems [25-27]. From the FFS variants available, we apply here the constrained branched growth (CBG) algorithm [28-31], as it has been shown to be efficient and to allow a cost-free estimation of committor probabilities ($p_B$) [31]. To estimate the nucleation free-energies, we have used a multiple-window Umbrella Sampling (US) method [29,30]. We use the number of particles in the largest (translationally ordered) solid-cluster, $n_{tr}$, as a reaction coordinate. To estimate $n_{tr}$, we adopted the order parameter introduced by ten Wolde *et al.* [5] to distinguish between liquid-like and solid-like particles as pertaining to the mesophases that the polyhedra form. Details on the calculation of $n_{tr}$ and transition rates are provided in [24].

The simulations were conducted in a cubic cell with periodic boundary conditions, with fixed number of particles, $N$ (=3375 for COs, 2500 for TOs, and 2916 for RDs), and constant reduced pressure $P^* = P\sigma^3/8$, where $\sigma$ is the diameter of the circumscribing sphere (specific to each particle geometry) and $\beta=1/k_BT$. $P^*$ values are chosen such that the degree of supersaturation (*DoSS*) approaches the range of values for which nucleation studies for HSs have been performed. The *DoSS* is calculated via [32]:

$$DoSS = \beta|\Delta\mu| \qquad (1)$$

where $|\Delta\mu|$ is the chemical potential difference between the metastable fluid and the stable solid at the simulated $P^*$, obtained using thermodynamic integration (see [24]). While several metrics of supersaturation can be used, $|\Delta\mu|$ embodies the true thermodynamic force driving the nucleation of the new phase. The values of $P^*$ and *DoSS* are given in Table 1. *DoSS* values > 0.55 were not explored since the liquid phase would become unstable. We performed FFS calculations at *DoSS* values of: 1) 0.3 and 0.38 for COs, 2) 0.3 and 0.42 for TOs, and 3) 0.27 and 0.36 for RDs, while we used a brute force approach similar to that used in Ref. [33] for the largest $DoSS$ for all shapes. We used kinetic MC (KMC) simulations to evolve trajectories from liquid to solid as in Ref. [34] and [35]; further simulation details are given in [24].

Table 1 shows the calculated nucleation rates for COs, TOs, and RDs. The nucleation rate, $I$ is given by $I = (T_{AB}V)^{-1}$, where $V$ is the volume of the system and $T_{AB}$ is mean first passage time from the liquid basin to the rotator basin. Rates are given in dimensionless form $I^* = I\sigma^5/D_0$, where $D_0$ is the free-particle diffusion constant (see Ref. [34,35] for details) and is assumed to be comparable for all four shapes. In Fig. 1(a), we show the free-energy profile over $n_{tr}$ obtained from US. In Table 1 and Fig. 1(b) we compare the nucleation rate and free-energy barrier height $\Delta G^*$ among HSs, COs, TOs, and RDs. As expected, $I^*$ increases and $\Delta G^*$ decreases with $DoSS$ for all shapes. RDs appear to have smaller $\Delta G^*$ and larger rate than the other two polyhedra. While RDs are space-filling like TOs and have an asphericity intermediate between TOs and COs, one feature unique to RDs is that all their 12 facets are identical (while both TOs and COs have two distinct facet shapes). It may be that such symmetry eliminates facet-to-facet mismatches between neighboring RD particles, hence facilitating their ordered assembly. Our results also show that for all three polyhedra $\Delta G^*$ is significantly smaller and the nucleation rate significantly larger than those for HSs. This is intriguing considering that all these hard-particle systems (HSs and COs in particular) could be thought of as cases where a similar translationally-ordered, rotationally-disordered phase is being nucleated. To find some clues to explain such disparity, we proceed to examine the microstructure of the nuclei using the configurations collected at different FFS interfaces.

First we note that, like in HSs, the growing nucleus in the polyhedra tends to have a spherical shape (see [24]). To characterize translational order, we calculate $f_\alpha$, the fraction of crystal structure type $\alpha$ in such a nucleus (see also [24] for details). Figure 2(a) shows the results of $f_\alpha$ as a function of $n_{tr}$ for COs, TOs, and RDs. Similar to HSs [30,32], the small clusters for the polyhedra have some bcc- and liquid-like signatures but unlike HSs [36], the most dominant structure is hcp with very small contribution of fcc. However, the very small structural dissimilarity between hcp and fcc lattices seems an unlikely major contributor to explain the lower nucleation free energy barriers in the polyhedra.

To quantify orientational order, we use parameters $P_4$ and $I_4$ [24,37] which capture any trace of the cubic orientation symmetry that COs, TOs, and RDs will fully attain in the crystalline state. In accord to previous observations [16,17], particles in the bulk rotator phases of these polyhedra near the isotropic-rotator coexistence pressure dynamically rotate (with liquid-like rotational diffusivity) in all possible directions, but do have a small amount of long-range orientational order. This is a reflection that particles are trapped in (translationally ordered) "cages" that are not perfectly spherical. We show in the *Supplemental Material* [24] that, for the COs in particular, the extent of long-range orientational order is small enough for the rotator phase to be considered orientationally disordered at any of the pressures investigated. As it turned out, however, this small long-range orientational order in the rotator phase of polyhedra does foreshadow the microscopic origin of their faster nucleation kinetics. To characterize orientational order in the growing nucleus, we calculate $P_4^{Cl}$, the cubatic-order parameter of the greatest cluster for different values of $n_{tr}$ [see Fig. 2(b)]. In all cases, $P_4^{Cl}$ stays in the narrow range from 0.08 to 0.11 which is a small value but still an order of a magnitude larger than the value of 0.011 for the bulk liquid phase of all three polyhedra, indicative of incomplete orientational disorder in the solid nuclei. $P_4^{Cl}$ is, however, smaller than the average $P_4$ for the corresponding stable bulk-rotator phases (see [24]). To probe whether a distinguishable orientationally ordered cluster exists and whether it is correlated with the translationally ordered nucleus, we define an order parameter, $n_{or}$, as the number of particles in the greatest cluster based on the rotationally invariant orientational order parameter, $I_4$ [37]. Akin to the translational order parameter, we define a local orientational order parameter $I_4(i)$ for particle $i$ (see [24]) and obtain $n_{or}$ with a procedure similar to that used to calculate $n_{tr}$. We define as $f_{overlap}$ the fraction of particles in $n_{or}$ that are present in $n_{tr}$. As shown in Fig. 2(b), $f_{overlap} \sim$ 10-20% at the pre-critical stages, increasing to $\sim$ 40% or higher at the critical stage for all cases. This shows that a significant degree of spatial coupling exists between orientational and translational order as the rotator-phase nucleates.

To further characterize the spatial coupling between orientational and translational order, we plot in Fig. 3(a) and 3(b) the normalized density distributions of the local translational order [$q_6(i)$] and local orientational order [$I_4(i)$] for the bulk isotropic and rotator phases of the polyhedra. In Fig. 3(a), we also show the $q_6(i)$ density distributions of the bulk isotropic and solid phases of HSs. All the distributions are plotted for a *DoSS* value near 0.3. The isotropic phase $q_6(i)$ distributions of COs, TOs, RDs, and HSs are almost indistinguishable and exhibit significant overlap with those for the corresponding solid phases. This shows that, like in HSs, the isotropic phase of polyhedra contains small regions of high local translational order. But unlike HSs, which lack any degree of freedom for orientational order, the local orientational order in the bulk isotropic phase for the polyhedra spans a broad range of values between 0.1 and 0.4, around an average value of $\langle I_4(i) \rangle \sim 0.2$. Importantly, these $\langle I_4(i) \rangle$ values are much larger than the average $I_4$ values for the whole isotropic phase ($\sim$0.02), indicative of spatially heterogeneous (short-range) orientational order. Note that COs show the highest amount of overlap between the isotropic phase and rotator phase $I_4(i)$, likely a reflection of their lower average orientational order of the bulk-rotator phase. Figure 3(d) shows that, similar to the regions of high local translational order [also present in HSs as shown in Fig. 3(c)], the regions of high local orientational order are small in size (correlation length) and appear scattered throughout the system. Although these pervading orientationally ordered and translationally ordered "hot spots" do not always coincide, it is their spatial coupling what likely facilitates both the birth [Fig. 3(e)] and growth [Fig. 3(f)] of the translationally-ordered embryo.

One possible metric of the coupling between local orientational and translational order is $n_{both}$ defined as the number of particles in $n_{or}$ that are present in $n_{tr}$ (i.e., a measure of the overlap between the largest orientationally-ordered cluster and the translationally-ordered nucleus). If such a coupling is favorable to nucleation, then configurations with a larger $n_{both}$ would tend to be associated with a larger probability to commit to the rotator phase ($p_B$ value) and, conversely, configurations with high $p_B$ values would tend to exhibit larger $n_{both}$ values. As shown in Fig. 4(a), the average and maximum values of $n_{both}$ indeed increase with $p_B$. The inset of Fig. 4(a) shows that with increasing $p_B$ values, not only the average value of $n_{tr}$ increases but also the average value of $n_{both}$. These correlations of $n_{both}$ with $n_{tr}$ and $p_B$ suggest that the nucleus would tend to grow in regions with stronger spatial coupling between localized fluctuations of orientational and translational order. Some direct evidence of the latter comes from visual examination of the critical cluster; Figure 4(b) shows one example of how the cluster tends to grow in regions where it previously had both local orientational and translational order (compare snapshots I and IV).

In summary, the rotator-phase nucleation barrier heights (and rates) for COs, TOs, and RDs were found to be significantly lower (larger) than those for HSs at comparable *DoSS*. Our analysis reveals a coupling between spatial fluctuations of orientational and translational order present in the isotropic phase of the polyhedra. This coupling can be seen as a positive feedback loop wherein the spontaneous local alignment of particles [16,17,38] (wherein neighbor particles tend to pack their flat facets parallel to each other) help steer particles toward positions with translational order; conversely, regions having established translational order (i.e., at the interface of the nucleus) in turn make fluctuations with high orientational order more prevalent. In the context of classical nucleation theory (see Supplementary Material [24]), this coupling in fluctuations (absent in hard spheres) would be expected to produce two opposing effects on the nucleation rate: (1) Decrease the solid-liquid interfacial tension $\gamma$ and hence the free-energy barrier $\Delta G^*$, and (2) reduce the attempt rate (prefactor) as particles approaching the nucleus may also be temporarily "trapped" by other regions with correlated order. Overall, given the general nature of this mechanism, we conjecture that it will also be at play for other polyhedral shapes that form rotator phases, and it could even be used to catalyze solid-nucleation of spheres by adding enough polyhedral particles as *initiators*.

**Acknowledgements**.
This work was supported by the U.S. National Science Foundation Grant No. CBET 1033349. The authors also thank U. Agarwal, Sai Pooja, and M. Kadhilkar for providing material for the codes used in this work, and to A. Gantapara and M. Dijkstra for providing additional data on their paper [18].

**Table 1:** Simulation data for the rotator-phase nucleation in COs, TOs and RDs. $\eta_{liquid}$ is the liquid phase packing fraction at pressure $P^*$. $\Delta G^*/k_B T$ is the free-energy barrier height and $n_{tr}^*$ is the critical cluster size.

| Shape | $P^*$ | $\eta_{liquid}$ | DoSS | $I\sigma^5/D_0$ | $\Delta G^*/k_B T$ | $n_{tr}^*$ |
|---|---|---|---|---|---|---|
| CO | 4.1 | 0.513 | 0.30 | $(1.9 \pm 0.5) \times 10^{-14}$ | $19.8 \pm 0.8$ | 177 |
| CO | 4.6 | 0.529 | 0.38 | $(2.0 \pm 0.6) \times 10^{-08}$ | $3.4 \pm 0.2$ | 55 |
| CO | 4.8 | 0.532 | 0.44 | $(1.0 \pm 0.5) \times 10^{-6}$ | $2.5 \pm 0.3$ | 31 |
| TO | 2.8 | 0.496 | 0.30 | $(1.0 \pm 0.5) \times 10^{-12}$ | $18.5 \pm 0.7$ | 165 |
| TO | 3.0 | 0.504 | 0.42 | $(5.8 \pm 0.8) \times 10^{-07}$ | $4.9 \pm 0.2$ | 59 |
| TO | 3.15 | 0.5085 | 0.52 | $(1.8 \pm 0.7) \times 10^{-05}$ | $1.7 \pm 0.1$ | 29 |
| RD | 4.32 | 0.5008 | 0.27 | $(3.2 \pm 1.3) \times 10^{-09}$ | $10.9 \pm 0.6$ | 117 |
| RD | 4.56 | 0.5072 | 0.36 | $(3.0 \pm 0.5) \times 10^{-06}$ | $3.2 \pm 0.1$ | 40 |
| RD | 4.64 | 0.511 | 0.42 | $(1.6 \pm 0.5) \times 10^{-05}$ | $1.9 \pm 0.1$ | 25 |

# Figures

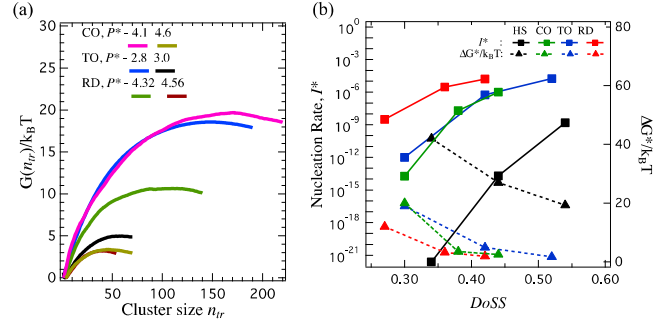

FIG. 1: a) Free-energy profile $G(n_{tr})/k_BT$ as a function of $n_{tr}$, and b) Free-energy barrier height $\Delta G^*/k_BT$ and Nucleation Rate, $I^*$ vs. $DoSS = \beta|\Delta\mu|$ for COs, TOs, and RDs. HS data from [36].

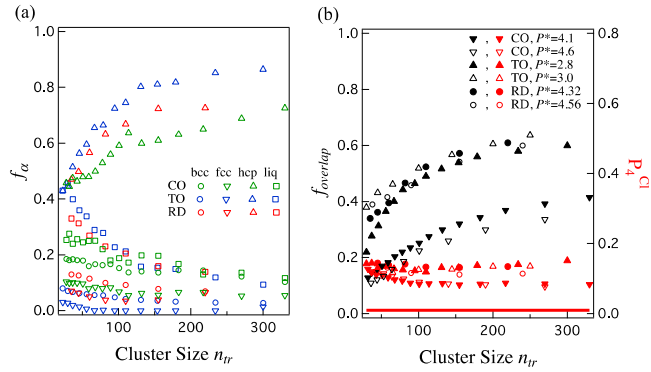

FIG. 2: Translational and orientational order in solid nuclei of COs, TOs, and RDs. a) Relative weight of the structural signatures for bcc, fcc, hcp and liquid-like ordering. b) Variation of $f_{overlap}$ and $P_4^{Cl}$ with $n_{tr}$. $P_4$ for the liquid phase is shown as a red line.

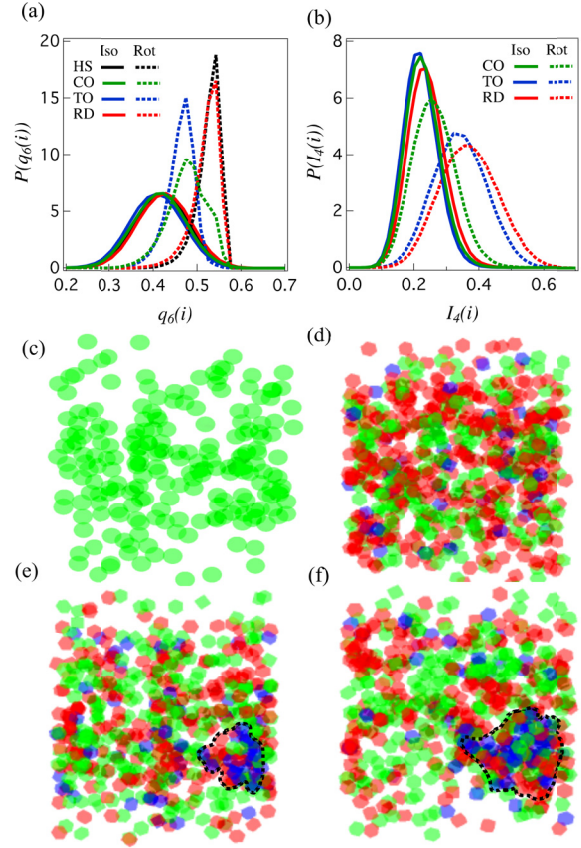

FIG. 3: Normalized Probability density distribution of (a) $q_6(i)$ for the isotropic and solid (rotator) phases of COs, TOs, RDs and HSs, and (b) $I_4(i)$ for the isotropic and rotator phases of COs, TOs and RDs. All configurations correspond to $DoSS \sim 0.3$. Snapshots for isotropic configurations of (c) HSs and (d) COs. Snapshots of CO configurations of (e) precritical-sized ($n_{tr} \sim 101$) and (f) critical-sized ($n_{tr} \sim 183$) nuclei (nucleus boundary marked by dashed black line). The snapshots only show selected particles: Translationally ordered particles (with $q_6(i) > 0.5$) are green, orientationally ordered particles (with $I_4(i) > 0.3$) are red, and translationally and orientationally ordered particles are blue.

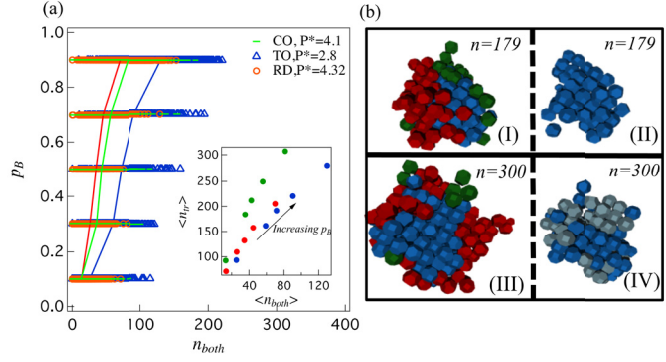

FIG. 4: (a) $p_B$ vs. $n_{both}$ for COs at $P^*$=4.1, TOs at $P^*$=2.8, and RDs at $P^*$=4.32. The inset shows the correlation between the average values of $\langle n_{both}\rangle$ and $\langle n_{tr}\rangle$ ($p_B$ values increase from the bottom-left to the top-right). (b) Snapshots of critical (I and II) and post-critical (III and IV) nuclei of the same trajectory for TOs at $P^*$=2.8. Blue TOs are both translationally and orientationally ordered, red TOs are translationally ordered, and green TOs are orientationally ordered. Snapshots II and IV only show blue TOs with snapshot IV also showing in gray the newly added particles.

#